\documentclass[aps,reprint,twocolumn,showpacs]{revtex4}
\usepackage{amsmath}
\usepackage{epsfig}
\usepackage{dcolumn}
\usepackage{bm}
\usepackage{amsfonts}
\usepackage{graphicx}
\usepackage{color}

\begin{document}

\title{Qubit transient dynamics at tunneling Fermi-edge singularity.}
\author{V.V. Ponomarenko$^1$ and I. A. Larkin$^{1,2}$}
\affiliation{$^1$Center of Physics, University of Minho, Campus
Gualtar, 4710-057 Braga, Portugal, $^2$ Institute of
Microelectronics Technology RAS, 142432 Chernogolovka, Russia}
\date{\today }

\begin{abstract}
We consider tunneling of spinless electrons from a single-channel
emitter into an empty collector through an interacting resonant
level of the quantum dot. When all Coulomb screening of sudden
charge variations of the dot during the tunneling is realized by
the emitter channel, the system is described with an exactly
solvable model of a dissipative qubit. We derive the corresponding
Bloch equation for its quantum evolution. We further use it to
specify the qubit transient dynamics towards its stationary
quantum state after a sudden change of the level position. We
demonstrate that the time-dependent tunneling current
characterizing this dynamics exhibits an oscillating behavior for
a wide range of the model parameters.
\end{abstract}

\pacs{73.40.Gk, 72.10.Fk, 73.63.Kv, 03.67Bg}

\maketitle

The generic response of conduction electrons in a metal to the
sudden appearance of a local perturbation results in the
Fermi-edge singularity (FES) initially predicted in \cite{1,2} and
also recently studied in the non-equilibrium systems \cite{AL}.
It was observed experimentally as a power-law singularity in X-ray absorption
spectra\cite{3,4}. Later, a possible occurrence of the FES in
transport of spinless electrons through a quantum dot (QD) was
considered \cite{5} in the regime when  a localized QD level is
below the Fermi level of the emitter in its proximity and the
collector is effectively empty (or in equivalent formulation through
the particle-hole symmetry). The Coulomb interaction with the charge
of the local level acts as a one-body scattering potential for the
electrons in the emitter. Then, in the perturbative approach
assuming a sufficiently small tunneling rate of the emitter, the
separate electron tunnelings from the emitter change the level
occupation and generate sudden changes of the scattering potential
leading to the FES in the I-V curves at the voltage threshold
corresponding to the resonance. Direct observation of these
perturbative results in experiments, however, is difficult because
of the finite life time of electrons in the localized state of the
QD, and in many experiments \cite{6,7,8,9} the FES's have been
identified simply by appearance of the threshold peaks in the I-V
dependence. According to the FES theory \cite{1,2} such peak could
occur when the exchange effect of the Coulomb interaction in the
tunneling channel exceeds the Anderson orthogonality catastrophe
effects in the screening channels and therefore signals formation
of an exciton electron-hole pair in the tunneling channel at the
QD. This pair can be considered as a two-level system
or qubit which undergoes dissipative dynamics. In the absence of
the collector tunneling and, if the Ohmic dissipation produced by
the emitter is weak enough, its dynamics are characterized
\cite{L,Sch} by the oscillating behavior of the level occupation,
which is beyond the perturbative description.

Therefore, in this work we study the qubit transient dynamics and
its manifestation in the collector tunneling current in a
simplified, but still realistic system described by a model
permitting an exact solution. It can be realized, in particular,
if the emitter is represented by a single edge-state in the integer
quantum Hall effect. In this system the Ohmic dissipation
produced by the emitter is absent and the qubit coherent
oscillations are only destroyed by the collector tunneling. Our
solution to this model will demonstrate when the observation of an
oscillatory behavior of the transient tunneling current is
possible and useful for further identification of FES in
tunneling experiments. We also find the stationary states of the
qubit to which the transient dynamics converge. We describe the
dependence of their Bloch vector on the
experimentally adjustable parameters of the setup and express their
entanglement entropy through the tunneling current. Being
controlled by the tunneling into the empty collector, the
stationary states in this model remain independent of temperature.

\emph{Model} - In the system we consider below, the tunneling
occurs from a single-channel emitter into an empty collector
through a single interacting resonant level of the QD located between
them. It is described with the Hamiltonian
$\mathcal{H}=\mathcal{H}_{res}+\mathcal{H}_C$ consisting of the
one-particle Hamiltonian of resonant tunneling of spinless
electrons and the Coulomb interaction between instant charge
variations of the dot and electrons in the emitter. The resonant
tunneling Hamiltonian takes the following form
\begin{equation}
\mathcal{H}_{res}=\epsilon _{d} d^+d+\sum_{a=e,c}\mathcal{H}_0[\psi_a]+
w_a(d^+\psi_a(0)+h.c.) \ ,  \label{hres}
\end{equation}
where the first term represents the resonant level of the dot, whose energy
is $\epsilon _{d}$. Electrons in the emitter (collector) are described with the chiral
Fermi fields $\psi_a(x),a=e(c)$, whose dynamics is governed by the
Hamiltonian $\mathcal{H}_0[\psi]=-i\!\! \int\! dx \psi^+(x) \partial_x
\psi(x) \ ( \hbar=1) $ with the Fermi level equal to zero or drawn to $%
-\infty$, respectively, and $w_a$ are the corresponding tunneling
amplitudes. The Coulomb interaction in the Hamiltonian $\mathcal{H}$ is
introduced as
\begin{equation}
\mathcal{H}_C=U_C \psi_e^+(0)\psi_e(0)(d^+d-1/2) \ .  \label{hc}
\end{equation}
Its strength parameter $U_C$ defines the scattering phase
variation $\delta$ for the emitter electrons passing by the dot
and therefore the screening charge in the emitter produced by a
sudden electron tunneling into the dot is equal to $\Delta
n=\delta/\pi \ \ (e=1)$ according to Friedel's sum rule. Below we
assume that the dot charge variations are completely screened by
the emitter tunneling channel and $\delta=-\pi$.

Next we implement bosonization and represent the emitter Fermi field as $%
\psi_e(x)=\sqrt{\frac{D}{2\pi} }\eta e^{i\phi(x)}$, where $\eta$ denotes an
auxiliary Majorana fermion and $D$ is the large Fermi energy of the emitter.
The chiral Bose field $\phi(x)$ satisfies $[\partial_x\phi(x),\phi(y)]=i2\pi%
\delta(x-y) $ and permits us to express
\begin{equation}
\mathcal{H}_0[\psi_e]=\int \frac{dx}{4 \pi} (\partial_x \phi)^2, \ \
\psi_e^+(0)\psi_e(0)=\frac{1}{2 \pi} \partial_x \phi(0) \ .  \label{hphi}
\end{equation}
Substituting these expressions into Eqs. (\ref{hres},\ref{hc}) we
find the alternative form for the Hamiltonian $\mathcal{H}$. By
applying the unitary transformation
$\mathcal{U}=\exp[i\phi(0)(d^+d-1/2)]$ to this form we come to the
Hamiltonian of the dissipative two-level system or qubit:
\begin{eqnarray}
\mathcal{H}_{Q}=\epsilon _{d} d^+d+\mathcal{H}_0+ w_c(\psi^+_c(0)e^{i\phi(0)}d+h.c.)
\notag \\
+\Delta \eta (d- d^+)+ (\frac{U_C}{2\pi}-1) \partial_x
\phi(0) (d^+d-\frac{1}{2}) \ ,  \label{hq} \\
\mathcal{H}_0=\mathcal{H}_0[\phi]+\mathcal{H}_0[\psi_c] \ ,
\notag
\end{eqnarray}
where $\Delta = \sqrt{\frac{D}{2\pi }}w_{e}$. This Hamiltonian is
further simplified. Since in bosonization technique the relation
\cite{schotte} between the scattering phase and the Coulomb
strength parameter is linear $\delta=-U_C/2$, the last term of the
Hamiltonian on the right-hand side of Eq. (\ref{hq}) vanishes and
also the bosonic exponents in the third term can be removed
because the time
dependent correlator of the collector electrons is $<\psi_c(t)\psi^+_c(0)>=%
\delta(t)$.

\emph{Bloch equations for the qubit evolution} - We use this
Hamiltonian to describe the dissipative evolution of the qubit
density matrix $\rho _{a,b}(t)$, where $a,b=0,1$ denote the empty
and filled levels, respectively. In the absence of the tunneling
into the collector at $w_{c}=0$, $\mathcal{H}_{Q}$ in Eq.
(\ref{hq}) transforms through the substitutions of $\eta
(d-d^{+})=\sigma _{1}$ and $ d^{+}d=(\sigma _{3}+1)/2$ ( $\sigma
_{1,3}$ are the corresponding Pauli matrices) into the Hamiltonian
$\mathcal{H}_{S}$ of a spin $1/2$ rotating in the magnetic field
$\mathbf{h}=(2\Delta , 0,\epsilon _{d})^{T}$ with the frequency
$\omega_0=\sqrt{4\Delta ^{2}+\epsilon _{d}^{2}}$ . Then the
evolution equation follows from
\begin{equation}
\partial _{t}\rho (t)=-i[\rho (t),\mathcal{H}_{S}]\ .  \label{rhos}
\end{equation}%
To incorporate in it the dissipation effect due to tunneling into
the empty collector we apply the diagrammatic perturbative
expansion of the S-matrix defined by the Hamiltonian (\ref{hq}) in
the tunneling amplitudes $w_{e,c}$ in the Keldysh technique. This
permits us to integrate out the collector Fermi field in the
following way. At an arbitrary time $t$ each diagram ascribes
indexes $a(t_{+})$ and $b(t_{-})$ of the qubit states to the upper
and lower branches of the time-loop Keldysh contour. This
corresponds to the qubit state characterized by the $\rho _{a,b}(t)$
element of the density matrix. The expansion in $w_{e}$ produces
two-leg vertices in each line, which change the line index into
the opposite one. Their effect on the density matrix evolution has
been already included in Eq. (\ref{rhos}). In addition, each line
with index $1$ acquires two-leg diagonal vertices produced by the
electronic correlators $<\psi _{c}(t_{\alpha })\psi
_{c}^{+}(t_{\alpha }^{\prime }) >,\ \alpha =\pm $. They result in
the additional contribution to the density matrix variation:
$\Delta \partial _{t}\rho _{10}(t)=-\Gamma \rho _{10}(t),\ \Delta
\partial _{t}\rho _{01}(t)=-\Gamma \rho _{01}(t),\ \Delta \partial
_{t}\rho _{11}(t)=-2\Gamma \rho _{11}(t),\ \Gamma =w_{c}^{2}/2$.
Then there are also vertical fermion lines from the upper branch
to the lower one due to the non-vanishing correlator $<\psi
_{c}(t_{-})\psi _{c}^{+}(t_{+}^{\prime })>$, which lead to the
variation $\Delta \partial _{t}\rho _{00}(t)=2\Gamma \rho
_{11}(t)$. Incorporating these additional terms into Eq.
(\ref{rhos}) and making use of
the density matrix representation $\rho (t)=[1+\sum_{l}a_{l}(t)\sigma _{l}]/2$%
, we find the evolution equation for the Bloch vector $\mathbf{a}(t)$ as
\begin{equation}
\partial _{t}\mathbf{a}(t)=M\cdot \mathbf{a}(t)+\mathbf{b}\ ,\ \mathbf{b}%
=[0,0,2\Gamma ]^{T}\ ,  \label{dadt}
\end{equation}%
where $M$ stands for the matrix:
\begin{equation}
M=\left(
\begin{array}{lll}
-\Gamma & -\epsilon _{d} & 0 \\
\epsilon _{d} & -\Gamma & -2\Delta \\
0 & 2\Delta & -2\Gamma%
\end{array}%
\right) \ .  \label{M}
\end{equation}%
Starting the evolution of the Bloch vector from its value $\mathbf{a}(0)$ at zero time,
we apply a Laplace transformation to Eq. (\ref{dadt}). Its inverse gives us this vector $%
\mathbf{a}(t)$ at positive time as following
\begin{equation}
\mathbf{a}(t)=\int_{C}\frac{dze^{zt}}{2\pi i}\left[ z-M\right] ^{-1}(\mathbf{%
a}(0)+M^{-1}\mathbf{b})-M^{-1}\mathbf{b}\ ,  \label{at}
\end{equation}%
where the integration contour $C$ coincides with the imaginary
axis shifting to the right far enough to have all poles of the
integral on its left side. These poles are defined by inversion of
the matrix $[z-M]$ and are equal to three roots of its determinant
$\det \left[ z-M\right] \equiv P(z)$, which is
\begin{equation}
P(z)=x^{3}+\Gamma x^{2}+(4\Delta ^{2}+\epsilon _{d}^{2})x+\Gamma
\epsilon _{d}^{2} \ ,\  x=z+\Gamma \ .  \label{p3}
\end{equation}%
Its roots $z_{l},\ l=\left\{ 0,1,2\right\} $ have their real parts negative.
Therefore, the stationary state of the qubit is characterized by the
Bloch vector:
\begin{equation}
\mathbf{a}(\infty )=-M^{-1}\mathbf{b}=\frac{[2\epsilon _{d}\Delta
,-2\Delta \Gamma ,(\epsilon _{d}^{2}+\Gamma ^{2})]^{T}}{\left(
\epsilon _{d}^{2}+\Gamma ^{2}+2\Delta ^{2}\right) }\ .
\label{ainfty}
\end{equation}%
In general, an instant tunneling current $I(t)$ into the empty
collector directly measures the diagonal matrix element of the
qubit density matrix \cite{us} through their relation
\begin{equation}
I(t)=2\Gamma \rho _{11}(t)=\Gamma \lbrack 1-a_{3}(t)]  \label{I-t}
\end{equation}%
It gives us the stationary tunneling current as $I_{0}=2\Gamma
\Delta^{2}/(2\Delta ^{2}+\Gamma ^{2}+\epsilon _{d}^{2})$. At
$\Gamma \gg \Delta $ this expression coincides with the
perturbative  results of \cite{5,lar}. Another important
characteristic is the qubit entanglement entropy
$S_e=-\mbox{tr}\{\rho \ln \rho\}$, which is just a function of the
Bloch vector length. The length of the stationary Bloch vector in
Eq. (\ref{ainfty}) is $|\mathbf{a}(\infty
)|=\sqrt{1-(I_0/\Gamma)^2}$. Therefore, measurement of the
tunneling current gives us also the entropy of the stationary
state of the qubit. This entropy changes from zero for the qubit
pure state of empty QD far from the resonance to its entanglement
maximum approaching $\ln 2$ at the resonance with an infinitely
small $\Gamma$.

The explicit form of the Laplace image $\widetilde{a_{3}}(z)$ in
Eq. (\ref{at}) is
\begin{equation}
\widetilde{a_{3}}(z)=\frac{a_{3}(\infty )}{z}+F(z),
\end{equation}%
where%
\begin{equation}
F(z)=\frac{1}{P(z)}\left( (\mathbf{f}(z) \cdot
\mathbf{a}(0))+f_{0}(z) \right)\ . \label{F}
\end{equation}%
The components of the vector $\mathbf{f}(z) $ are
$\ f_{1}(z)=2\epsilon _{d}\Delta ,$ $f_{2}(z)=2x\Delta $ , $%
f_{3}(z)=x^{2}+\epsilon _{d}^{2}$ and $f_{0}(z)=-(\mathbf{f}(z)
\cdot \mathbf{a}(\infty))$ is equal to
\begin{equation}
\ f_{0}(z)=-\frac{\epsilon _{d}^{4}+\left( x^{2}+\Gamma ^{2}+4\Delta
^{2}\right) \epsilon _{d}^{2}+x\Gamma \left( x\Gamma -4\Delta ^{2}\right) }{%
\left( \Gamma ^{2}+2\Delta ^{2}+\epsilon _{d}^{2}\right) } \ .
\end{equation}%

The inverse Laplace transform (\ref{at}) results in
\begin{equation}
a_{3}(t)=a_{3}(\infty )+\sum\limits_{l=0}^{2}r_{l}\cdot \exp [z_{l}t]
\label{at3}
\end{equation}%
where $z_{l}$ are the poles of $F(z)$ and $r_{l}$ are their
corresponding residues. In order to find these poles we bring the
cubic equation (\ref{p3}) to its standard form \cite{Jacobson}:
\begin{equation}
y^{3}+3Qy-2R=0 \   \label{y^3}
\end{equation}%
by applying the following notations $z=(y-4\Gamma)/3$ and
\begin{equation}
Q=12\Delta ^{2}-\Gamma^2 +3\epsilon _{d}^{2} ,\ \ \ R=\left(
18\Delta ^{2}-9\epsilon _{d}^{2}-\Gamma^2 \right)\Gamma \ .
\end{equation}%
\begin{figure}[b]
\centering \includegraphics{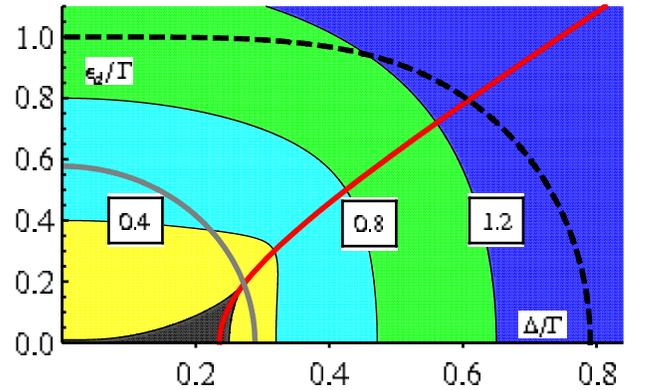}
\caption{ Contour plot of the positive imaginary part of the dimensionless root
Im$[y_{1}]/\Gamma=%
\frac{\protect\sqrt{3}}{2\Gamma}(S-T)$. The black area corresponds to the
region where all three roots are real. The red line corresponds to
$R=0$ and the gray line to $Q=0$. The black dashed curve shows
$Im[z_1]=-Re[z_1] $.} \label{fig:Im}
\end{figure}
\begin{figure}[b]
\centering \includegraphics{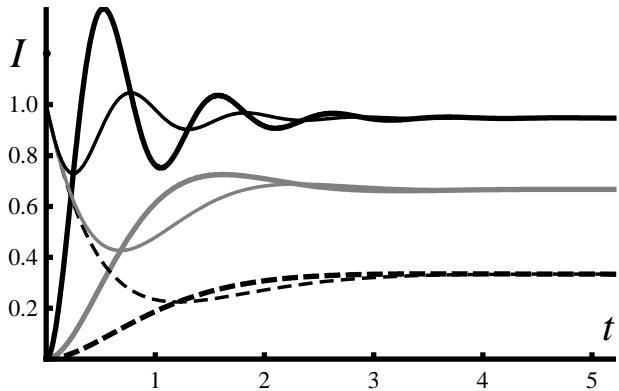} \caption{Plot of the
current $I(t)$ in Eq. \ref{ItEdeq0} when $\Gamma=1$ and
$\epsilon_{d}=0$. The black line corresponds to the parameter
$\Delta=3$, the gray line to $\Delta=1$, and the black dashed line to
$\Delta=0.5$. Thick lines correspond to the initially empty QD, thin
lines to the evolution starting from the zero Bloch vector.}
\label{fig:I01}
\end{figure}
The three roots are
\begin{equation}
y_{l}=e^{2/3\pi il}S+e^{-2/3\pi il}T \ ,  \label{y-roots}
\end{equation}%
where $l=0,~1,~2$ and%
\begin{equation}
S=\left( R+\sqrt{Q^{3}+R^{2}}\right)^{1/3}\text{ and }
T=-\frac{Q}{S} \ .
\end{equation}%
Here the function $\ Z^{1/3}$ of the complex variable $Z$ is determined in the
conventional way with the cut $Z\in \{-\infty ,0\}$. If the discriminant is
positive: $Q^{3}+R^{2}>0$, $S$ and $T$ are real positive and negative,
respectively. Therefore, the root $y_0$ is real and the two others $y_{1,2}$
are complex conjugates of each other. In the case of $Q^{3}+R^{2}<0$, $S$ and $T$ are also
complex conjugate. Hence, all three roots are real. In this case the
oscillatory behavior does not occur. This parametric area of triangular
form is depicted as black in Fig. 1. Its three vertices have coordinates
(0,0), (1/4,0) and ($\sqrt{2/27},\sqrt{1/27}$).

\emph{Oscillatory transient current} -
With $Q^{3}+R^{2}>0$ we find from Eqs. (\ref{I-t}) and (\ref{at3})
that
\begin{align}
I(t)& =I_{0}-\Gamma \left\{r_{0}\cdot e^{-G_0 t}+2 \text{Re}\left[
r_{1}\cdot
e^{-(G_1-i\omega )t} \right] \right\} , \notag \\
G_0&=\frac{4}{3}\Gamma -\gamma_1 \, , \, G_1=\frac{4}{3}\Gamma
+\frac{\gamma _{1}}{2} .  \label{ItQgt0}
\end{align}
Here $\gamma _{1}=\frac{1}{3}(S+T)$ and the second term in Eq. (\ref{ItQgt0} )
describes decaying
current oscillations of the frequency $\omega =%
\frac{\sqrt{3}}{6}(S-T)$. Note that the signs of $\gamma_1$ and
$R$ coincide. Therefore, above the line $R=0$ in Fig.\ref{fig:Im}
$\gamma_1$ is negative and the first term of the current in Eq.
(\ref{ItQgt0}) vanishes more quickly than the amplitude of the second-term
oscillations. Below this line $\gamma_1$ is positive and the
amplitude of the oscillations vanishes more quickly than the first
term. By differentiating Eq. (\ref{ItQgt0}) we find the condition for
the extrema of the current vs. time as
\begin{equation}
-\frac{r_{0} e^{\frac{3}{2} \gamma_{1}t}}{2  |r_{1}|}=
\frac{\sqrt{G_1^2+\omega^2}}{G_0}
\sin(\omega t+\varphi_r+\chi) , \label{extreme}
\end{equation}%
where $\varphi_r$ is the phase of $r_1$ and $\chi=\arctan(G_1/\omega)$.
In the parametric area of $R<0$ this equation shows
that the current is an infinitely oscillating function of time, while for $R>0$
the current will have a finite number of oscillations only if
$r_0/(2 |r_1|) \exp(3 \pi \gamma_1/(2 \omega))$ is less than the coefficient in front of the
sine function on the right-hand side of Eq. (\ref{extreme}). This condition is not very
restrictive and can be circumvented in general.
Indeed, contrary to the frequencies and the amplitude decay rates, the residues $r_{0,1}$
of the function $F(z)$ in Eq. (\ref{F}) depend
on the choice of the initial condition $\mathbf{a}(0)$ for the Bloch vector.
We can choose the initial condition by varying $\epsilon _{d}$ and $\Gamma $ to bring the
qubit into any desirable stationary state within the time of $\sim
1/\Gamma $ and further use this state as an initial condition to the
new transient evolution after abrupt change of these parameters.
In particular, by tuning $(\mathbf{f}(z_0) \cdot
\mathbf{a}(0))=(\mathbf{f}(z_0) \cdot \mathbf{a}(\infty))$ we make
$r_0$ vanish. Then, as follows from Eq. (\ref{extreme})
the transient current is always oscillating outside of the black area
in Fig. \ref{fig:Im}, but the direct visibility of these oscillations imposes
a stronger condition that $\omega > G_1$ as illustrated below. The border of this area
is marked by the black dashed line in Fig. \ref{fig:Im}.

At the resonance $(\epsilon _{d}=0)$ the root $z_{0}$ of $P(z)$ in Eq. (\ref%
{p3}) comes as $z_{0}=-\Gamma $. The vector $\mathbf{f}(z_0)$ is
zero and so is $r_0$ for any initial condition. The current is
infinitely oscillating with the frequency $\omega_r=\sqrt{4\Delta
^{2}-\Gamma ^{2}/4}$ if $\Delta >1/16$ and the decay rate of the
oscillations amplitude is $G_1=(3/2)\Gamma $. The general
expression for the residue $r_1$ dependence on the initial
conditions in Eq. (\ref{ItQgt0}) can be found as
\begin{align}
r_{1}& =\left( 1+i\frac{\Gamma }{2\omega_r}\right) a_{3}(0)-i\frac{%
2\Delta }{\omega_r}a_{2}(0)  \notag \\
& -\Gamma \frac{2\omega_r\Gamma +i(\Gamma ^{2}+8\Delta
^{2})}{4\omega_r(\Gamma ^{2}+2\Delta ^{2})} \, .  \label{r1res}
\end{align}
We consider first the experimentally feasible case of the qubit evolution
from the initial state corresponding to the empty QD.

The empty
QD may be prepared by application of the bias voltage to the emitter
to make  $\epsilon _{d}\gg \Delta, \Gamma .$ Then the state of the qubit, as follows from
Eq. (\ref{ainfty}), is defined by $%
a_{1}(0)=a_{2}(0)=0$ and $a_{3}(0)=1$ and corresponds to the zero
tunneling current. In the resonance case the substitution of Eq.
(\ref{r1res}) with these initial conditions into Eq.
(\ref{ItQgt0}) produces a simple formula for the current
oscillations
\begin{equation}
I(t)=I_{0}\left( 1-\text{Re}\left[ \frac{\omega_r-3i\Gamma /2}{\omega_r%
}\cdot \exp [-\frac{3}{2}\Gamma t+i\omega_r t]\right] \right) .
\label{ItEdeq0}
\end{equation}
This current dependence on time is depicted in Fig. \ref{fig:I01}
by thick lines for three different values of $\Delta$, which
correspond to $\omega_r=0.8667$ in the case of the dashed line and
$\omega_r=1.94$ and $\omega_r=5.98$ for the gray and black solid
lines, respectively. From Eq. (\ref{extreme}) we find the extrema
of the current in Eq. (\ref{ItEdeq0} ) to be exactly at
$t_n=n\pi/\omega_r$. Although the current is always an oscillating
function these oscillations become visible first for the gray line
in accordance with our criterion $\omega_r \ge G_1$.

In Fig.\ref{fig:I01} we also draw three thin lines of the current
dependence on time for the same three values of the rate $\Delta$
in the case of the qubit evolution with the initial condition of
the zero Bloch vector $\mathbf{a}(0)=0$. The current starts from
the finite value $I(0)=\Gamma$. This makes the oscillations of all
three lines more visible as their first extremes are located at
approximately twice smaller times. This initial state of the qubit
can be prepared, in particular, by making the collector tunneling
rate $\Gamma$ infinitely small at the resonance. It also could be
reached through thermodynamical equilibration of the qubit with
the high temperature emitter in the absence of tunneling between
QD and the collector due to by some slow dissipation processes
unaccounted for in our model. We have performed our calculations
in dimensionless units with $\hbar=1$ and $e=1$. In the experiment
\cite{lar, lar1} the collector tunneling rate is $\Gamma \approx
0.1 meV$ and the parameter $\Delta \approx 0.016 meV$. This
corresponds to the stationary current $I_0\approx 1.2 nA$. To
observe the regime of oscillations as shown in Fig. 2 (gray line)
one can take a heterostructure with $\Gamma=\Delta$. For example,
with $\Gamma=\Delta=0.01meV$ the stationary current is $I_0=1.62
nA$. The unit of time $t$ in Fig. 2 for this value of $\Gamma$ is
equal to $65.8 ps$.

In conclusion, the spinless electrons tunneling through an
interacting resonant level of a QD into an empty collector has
been studied in the especially simple, but realistic system, in
which all sudden variations in charge of the QD are effectively
screened by a single tunneling channel of the emitter. Making use
of the exact solution to this model, we have demonstrated  that
the FES in the tunneling current dependence on voltage should  be
accompanied by oscillations of the time-dependent transient
tunneling current in a wide range of the model parameters. In
particular, they occur if the emitter tunneling coupling $\Delta$
or the absolute value of the resonant level energy $|\epsilon_d|$
are large enough in comparison to the collector tunneling rate
$\Gamma$ and either $\Delta>\Gamma/4$ or
$\epsilon_d^2>\Gamma^2/27$ holds. These oscillations result from
the emergence of the qubit composed of electron-hole pair at the
QD and its coherent dynamics. The qubit can be manipulated by
changing voltage and the tunneling rates in the system.
\bigskip

\emph{Acknowledgment} - The work was supported by the Foundation
for Science and Technology of Portugal and by the European Union
Seventh Framework Programme (FP7/2007-2013) under grant agreement
n$^{\mathrm{o}}$ PCOFUND-GA-2009- 246542 and Research Fellowship
SFRH/BI/52154/2013. .

\end{document}